\documentclass[pra,aps,twocolumn]{revtex4-1}
\usepackage{graphicx}           
\usepackage{bm}                 
\usepackage{amsmath}            
\usepackage{amssymb}
\usepackage{verbatim}           
\usepackage{amsthm}             
\theoremstyle{plain}            

\def\bra#1{{\langle#1|}}
\def\ket#1{{|#1\rangle}}
\def\bracket#1#2{{\langle#1|#2\rangle}}

\def\expect#1{{\langle#1\rangle}}

\def\tr{{\rm Tr}}

\begin{document}

\title{Decoherence by Internal Degrees of Freedom}
\author{Todd A. \surname{Brun}}\email{tbrun@usc.edu}
\author{Leonard \surname{Mlodinow}}\email{lmlodinow@gmail.com}
\affiliation{Center for Quantum Information Science and Technology, University of Southern California, Los Angeles, California}

\date{\today}

\begin{abstract}
We consider a composite system consisting of coupled particles, and investigate decoherence due to coupling of the center-of-mass degree of freedom with the internal degrees of freedom.  For a simple model of two bound particles, we show that in general such a decoherence effect exists, and leads to suppression of interference between different paths of the center-of-mass.  For the special case of two harmonically-bound particles moving in an external potential in one dimension, we show that the coupling between the center-of-mass and internal degrees of freedom can be approximated as parametric driving, and that nontrivial coupling depends on the second derivative of the external potential.  We find a partial solution to this parametric driving problem.  For a simple interference experiment, consisting of two wave packets scattering off of a square well, we perform numerical simulations and show a close connection between suppression of interference and entanglement between the center-of-mass and internal degrees of freedom.  We also propose a {\it measure of compositeness} which quantifies the extent to which a composite system cannot be approximated as a single, indivisible particle.  We numerically calculate this quantity for our square well example system.
\end{abstract}

\pacs{03.65.Yz 03.65.Ta 03.67.-a 03.67.Mn}

\maketitle

\section{Introduction}

Though the fundamental components of all physical systems follow the laws of quantum mechanics, the objects in our everyday reality obey Newton's laws. Clearly, in composite objects, the quantum interactions of numerous particles somehow conspire to produce a ``whole'' that obeys Newtonian physics. Unfortunately, quantum theory does not provide a natural or obvious framework for the derivation of that classical reality, or for indicating the border between quantum and classical behavior.

In the early days of quantum theory the demarcation between quantum and classical behavior was often placed, at least approximately, at the boundary between the microscopic and macroscopic. But this couldn't be the ultimate answer, for no object is truly macroscopic in the sense that it is a single particle of macroscopic dimension and weight. Instead, to call something macroscopic is to approximate it by ignoring the internal states of its microscopic components, thereby essentially reducing the problem to center-of-mass variables.

In the past few decades, the failures of the macro/micro distinction have also been highlighted by new experimental results. For example, a nanometer-scale superconducting electrode connected to a reservoir via a Josephson junction constitutes a macroscopic two-level electronic system in which the two charge states can be coherently superposed \cite{Joyez94,Bouchiat98,Nakamura99}. More recently, nanomechanical resonators coupled to electrical circuits---systems of billions of atoms---have been measured to be oscillating in their quantum ground state \cite{Chan11}. But the grandest demand for a seamless approach to the quantum/classical divide comes from cosmology: since the universe has no place for an external observer, a theory of quantum cosmology must include classical physics as a property that emerges from purely quantum laws \cite{Hartle83}.

The mechanism by which the classical world emerges from a set of quantum particles is called decoherence. One way systems decohere is through interactions with an external system, the environment, which typically consists of a large number of degrees of freedom, and is not subject to measurement. The coupling means that the state of the environment will be sensitive to the state of the system, and will in a sense, ``observe'' the system, leading to a destruction of coherence. Mathematically, this occurs because our lack of knowledge about the environment requires that we trace over the environmental degrees of freedom, leading to an evolution of the observed system in which the density matrix is reduced to a classical probability distribution. There are many environmental mechanisms that can cause this. For example, the molecules in a beam can collide with lighter particles \cite{Hornberger03,Hornberger04}, or interact with thermal radiation \cite{Hackermuller04}.

While much work has been done to examine the decoherence of systems due to external factors (see, for example, \cite{Bray82,ZurekPhysToday,WallsMilburn94,Buzek95}) as cosmologists have pointed out, we need not turn to external systems as the cause of the emergence of classical behavior. Such behavior can arise from the make-up of the object itself---that is, from the object's own internal states, which can act essentially as an ``external environment'' that travels with the object, allowing the body's center-of-mass motion to be approximated by the classical equations of motion \cite{Hillery05,Nikolic15}.

In this paper we investigate the mechanism for this transition to classical behavior through the influence-functional methods of Feynman and Vernon \cite{Feynman63}. That method is tailor-made for this analysis because it allows us to integrate out the internal degrees of freedom, leaving an expression for the center-of-mass coordinates alone. We will then illustrate the effect by considering a specific scattering problem. We find that when the scattered particles possess internal states, as the entanglement between internal state and environment grows, the interference diminishes.

\section{The influence functional}

For simplicity of analysis, we will consider a composite system with only a single internal degree of freedom.  Obviously, more complicated systems with many internal degrees of freedom will likely exhibit much stronger decoherence effects, but analyzing even a single internal degree of freedom is both revealing and fairly challenging.

\subsection{Propagator}

Consider a two-particle system in an external potential $V$, and subject to an internal potential $U(x_1,x_2)$. Assume both particles have mass $m$.  We can change variables to separate the center-of-mass and internal degrees of freedom:
\begin{eqnarray}
Y &\equiv& \frac{1}{2}\left( x_1 + x_2 \right) , \nonumber\\
y &\equiv& \frac{1}{2}\left( x_1 - x_2 \right) ,
\end{eqnarray}
or inverting the relationship,
\[
x_{1,2} = Y \pm y .
\]

In terms of these new variables, the kinetic energy is
\[
\frac{1}{2} m \left(\dot{x}_1^2 + \dot{x}_2^2\right) = \frac{1}{2} M \left(\dot{Y}^2 + \dot{y}^2\right) ,
\]
where $M = 2m$ is the mass of the composite system. Assuming that $U(x_1,x_2)$ has the form $U(x_1,x_2) = U(y)$, the potential energy is
\[
V(x_1) + V(x_2) + U(x_1,x_2) = V(Y+y) + V(Y-y) + U(y) .
\]
The Lagrangian in terms of $Y$ and $y$ is thus
\begin{equation}
L(Y,y,\dot{Y},\dot{y}) = \frac{1}{2} M \left(\dot{Y}^2 + \dot{y}^2\right) - V(Y+y) - V(Y-y) - U(y) .
\end{equation}

The propagator for this system is given by the path integral
\begin{eqnarray}
&& K(Y_f,y_f,t_f; Y_i,y_i,t_i) \nonumber\\
&=& \int\int e^{ \frac{i}{\hbar}\int_{t_i}^{t_f} L(Y(t),y(t),\dot{Y}(t),\dot{y}(t)) dt }
\mathcal{D}y\, \mathcal{D}Y \nonumber\\
&=& \int e^{ \frac{iM}{2\hbar} \int_{t_i}^{t_f} \dot{Y}^2 dt }
  \tilde{K}[Y(t); y_f, t_t; y_i, t_i) \mathcal{D}Y,
\label{eq:propagator}
\end{eqnarray}
where $\tilde{K}[Y(t); y_f, t_t; y_i, t_i)$ is
\begin{eqnarray}
&& \tilde{K}[Y(t); y_f, t_t; y_i, t_i) \nonumber\\
&=& \int \exp \frac{i}{\hbar} \biggl\{  \int_{t_i}^{t_f}
  \biggl( \frac{M}{2} \dot{y}^2(t) - V(Y(t)+y(t)) \nonumber\\
 && - V(Y(t)-y(t)) - U(y(t)) \biggr)  dt \biggr\} \mathcal{D}y .
\label{eq:internalPropagator}
\end{eqnarray}
If $\tilde{K}[Y(t); y_f, t_t; y_i, t_i) = e^{ \frac{i}{\hbar} \int_{t_i}^{t_f}
\left(- 2V(Y(t))) \right) dt }$ the composite-particle propagator in (\ref{eq:propagator}) would represent the propagator for a single particle of mass $M = 2m$ and ``charge'' 2, i.e., moving in a potential $2V(Y)$. This is a good approximation when the contributions from the path integral in (\ref{eq:internalPropagator}) are dominated by the path $y(t) = \dot{y}(t) = 0$. In general, though this may not hold, the paths that are important in (\ref{eq:internalPropagator}) are nevertheless of order $L$, the ``size'' of the composite system. If the external potential, $V(Y)$, varies slowly over such distances, this suggests approximating (\ref{eq:internalPropagator}) employing a Taylor expansion around $Y(t)$:
\[
V(Y+y) + V(Y-y) = 2V(Y) + V''(Y) y^2  + O(y^4).
\]
The $2V(Y)$ term is just a potential acting on the center-of-mass degree of freedom, and can be absorbed into the action for the center-of-mass.  The second term has the form of parametric driving on the internal degree of freedom. 
This gives the following approximation for the propagator in (\ref{eq:propagator}):
\begin{eqnarray}
\label{eq:propagatorThree}
&& K(Y_f,y_f,t_f; Y_i,y_i,t_i) \nonumber\\
&=& \int\mathcal{D}Y
  e^{\frac{i}{\hbar}\int_{t_i}^{t_f} \left( \frac{M}{2} \dot{Y}^2(t) - 2V(Y(t)) \right) dt} \\
&&\times  \int\mathcal{D}y
  e^{\frac{i}{\hbar}\int_{t_i}^{t_f} \left( \frac{M}{2} \dot{y}^2(t) - V''(Y(t))y^2(t)-U(y(t)) \right) dt} . \nonumber
\end{eqnarray}
We will make this approximation in the following section, in which we consider the  effective transition probabilities for center-of-mass states, assuming that the final state of the internal degree of freedom goes unobserved. Note that, since the harmonic force driving the internal state variable depends on the path in the center-of-mass path integral, there is no equivalent approximation in the Hamiltonian formalism.

\subsection{Transition Probability}

From the propagator we can derive the probability for the degrees of freedom to make a transition from an initial product wave function $\phi(Y_i)\psi_i(y_i)$ to a final product wave function $\chi(Y_f)\psi_f(y_f)$:
\begin{widetext}
\begin{equation}
p\biggl(\phi(Y_i)\psi_i(y_i) \rightarrow \chi(Y_f)\psi_f(y_f)\biggr) =
\Biggl| \int\int\int\int  \chi^*(Y_f)\psi^*_f(y_f) K(Y_f,y_f,t_f; Y_i,y_i,t_i) \phi(Y_i)\psi_i(y_i) dY_f dy_f dY_i dy_i \Biggr|^2 .
\end{equation}
\end{widetext}
We fix the initial state of the internal degree of freedom and obtain an effective transition probability for the center-of-mass state alone by summing over (i.e., tracing out) the final state of the internal degree of freedom:
\begin{eqnarray}
&& p\biggl(\phi(Y_i) \rightarrow \chi(Y_f)\biggl| \psi_i(y_i)\biggr) = \\
&& \int\int\int\int dY_f dY'_f dY_i dY'_i \chi(Y_f)\chi^*(Y'_f)  \phi(Y_i)\phi^*(Y'_i) \nonumber\\
&& \times  \int\int \mathcal{D}Y\mathcal{D}Y' e^{\frac{iM}{2\hbar}\int_{t_i}^{t_f}
  \left( \dot{Y}^2(t) - \dot{Y'}^2(t) \right) dt} \bar{F}_\psi[Y(t),Y'(t)] , \nonumber
\label{eq:transitionProb}
\end{eqnarray}
where $\bar{F}_\psi[Y(t),Y'(t)]$ is the {\it influence functional}:
\begin{eqnarray}
\bar{F}_\psi[Y(t),Y'(t)] &=& \int\int\int dy_f dy_i dy'_i \psi^*_i(y'_i) \psi_i(y_i) \nonumber\\
&& \times \tilde{K}^*[Y'(t); y_f, t_t; y'_i, t_i) \\
&& \times \tilde{K}[Y(t); y_f, t_t; y_i, t_i) . \nonumber
\label{eq:influenceFunctional}
\end{eqnarray}
This integral has an interesting interpretation.  The center-of-mass trajectories in the path integral for the propagator of the full system (\ref{eq:propagator}) appear here as fixed functions of time, $Y(t)$ and $Y'(t)$. If we consider these to be classical driving functions, then we can define two wave functions
\begin{eqnarray}
\tilde{\psi}_Y(y_f) &=& \int dy_i \tilde{K}[Y(t); y_f, t_f; y_i, t_i)\psi_i(y_i) , \nonumber\\
\tilde{\psi}_{Y'}(y_f) &=& \int dy'_i \tilde{K}[Y'(t); y_f, t_f; y'_i, t_i)\psi_i(y'_i) .
\end{eqnarray}
These are the wave functions of the internal degree of freedom that would result if that degree of freedom began in the state $\psi_i$, and was then ``classically driven'' by a potential that depends on the center-of-mass trajectories $Y(t)$ and $Y'(t)$, respectively, namely:   
\[
V(Y(t)+y) + V(Y(t)-y) + U(y) 
\]
and
\[
V(Y'(t)+y) + V(Y'(t)-y) + U(y).
\]
The influence functional is the scalar product of those two wave functions at time $t_f$:
\begin{equation}
\bar{F}_\psi[Y(t),Y'(t)] = \bracket{\tilde{\psi}_Y}{\tilde{\psi}_{Y'}} .
\end{equation}
Though we have a simple physical interpretation, evaluating the path integrals in (\ref{eq:transitionProb}) for general potentials is very difficult.  But we can make more progress, and see a simpler physical interpretation, if we use the approximation we alluded to above:
\[
V(Y+y) + V(Y-y) = 2V(Y) + y^2 V''(Y) + O(y^4).
\]
We can now more cleanly separate the terms that depend only on $Y(t)$ and $Y'(t)$ from those that depend on on $y(t)$ and $y'(t)$, giving us the following approximate expression for the influence functional:
\begin{eqnarray}
\label{eq:propagatorTwo}
&& \bar{F}_\psi[Y(t),Y'(t)] \nonumber\\
&=& e^{-\frac{2i}{\hbar} \int_{t_i}^{t_f} (V(Y(t))-V(Y'(t))) dt} \nonumber\\
&& \times \int dy_f \left[ \int dy'_i K^*[Y'(t); y_f, t_f;y'_i, t_i) \psi^*_i(y'_i) \right] \nonumber\\
&& \times \left[ \int dy_i K[Y(t); y_f, t_f;y_i, t_i) \psi_i(y_i) \right] \nonumber\\
&=& e^{-\frac{2i}{\hbar} \int_{t_i}^{t_f} (V(Y(t))-V(Y'(t))) dt}\bracket{{\psi}_Y}{{\psi}_{Y'}} ,
\end{eqnarray}
Here, $K[Y(t); y_f, t_f;y_i, t_i)$ is the propagator corresponding to the time-dependent (and center-of-mass-path-dependent) Hamiltonian
\[
H_Y(t) = \frac{p_y^2}{2M} + U(y) + V''(Y(t))y^2 ,
\]
and $\psi_Y(y_f)$ is the wave function of the internal degree of freedom at time $t_f$ after evolving via that Hamiltonian from the initial wave function $\psi_i(y_i)$. 
Employing (\ref{eq:propagatorTwo}) we obtain the following approximation for (\ref{eq:transitionProb}):
\begin{widetext}
\begin{eqnarray}
p\biggl(\phi(Y_i) \rightarrow \chi(Y_f)\biggl| \psi_i(y_i)\biggr) &=&
\int\int\int\int dY_f dY'_f dY_i dY'_i \chi(Y_f)\chi^*(Y'_f)  \phi(Y_i)\phi^*(Y'_i) \nonumber\\
&& \times  \int\int \mathcal{D}Y\mathcal{D}Y' e^{\frac{iM}{2\hbar}\int_{t_i}^{t_f}
  \left( \dot{Y}^2(t) - \dot{Y'}^2(t) \right) dt} e^{-\frac{2i}{\hbar} \int_{t_i}^{t_f} (V(Y(t))-V(Y'(t))) dt}\bracket{{\psi}_Y}{{\psi}_{Y'}}.
\label{eq:transitionProbApprox}
\end{eqnarray}
\end{widetext}
This expression shows clearly that the effect of the internal states is to modulate the integrand for the center-of-mass motion through the factor $\bracket{{\psi}_Y}{{\psi}_{Y'}}$. When $\bracket{{\psi}_Y}{{\psi}_{Y'}} = 1$, the compound particle acts as if it is not composite, but rather a single particle with mass equal to the sum of the component masses. The degree to which $\bracket{{\psi}_Y}{{\psi}_{Y'}}$ deviates from unity determines the "degree of compositeness" of the compound particle. 

We expect decoherence effects---such as the suppression of interference---when $|\bracket{{\psi}_Y}{{\psi}_{Y'}}| < 1$ for $Y(t) \ne Y'(t)$.  This will depend on both the internal potential $U(y)$ and the coupling term $V''(Y(t))$.  If $V(Y)$ is a quadratic potential---for instance, a harmonic oscillator---then this term will be constant and there will be no coupling between the internal and external degrees of freedom.  Therefore decoherence should only occur if the external potential $V(Y)$ is not harmonic.  However, even in that case, the effects will also depend on the {\it internal} potential, as well as the state of the internal degree of freedom. We now specialize to the case where the internal potential is harmonic.

\section{Degree of compositeness}
\label{sec:compositeness}

As mentioned after Eq.~(\ref{eq:transitionProbApprox}), a natural question arises as to what degree a compound system can be treated as a single particle, rather than a composite system.  It would be useful to have a measure of ``compositeness,'' which would indicate to what degree we can ignore the internal degrees of freedom of the system.  To put this another way, we would like to ask to what extent we can describe the behavior of the center-of-mass degree of freedom while neglecting the internal degree (or degrees) of freedom.

Consider a composite system in an initial product state
\[
\ket{\Psi_0} = \ket{\phi_0}_Y \otimes \ket{\psi_0}_y .
\]
The Hamiltonian of the joint system is given by 
\begin{eqnarray}
H &=& \frac{p_Y^2}{2m} + \frac{p_y^2}{2m} + V(Y+y) + V(Y-y) + U(y) \nonumber\\
&\approx& \frac{p_Y^2}{2m} + \frac{p_y^2}{2m} + 2V(Y) + V''(Y)y^2 + U(y) .
\end{eqnarray}
The term $V''(Y)y^2$ couples the internal and center-of-mass degrees of freedom.  How much does this influence the evolution of the center-of-mass?  This will in general depend both on the particular system (the choice of potentials $V(Y)$ and $U(y)$) and the states of the internal and external degrees of freedom.

We can define a ``decoupled'' Hamiltonian
\begin{equation}
H_d = \frac{p_Y^2}{2m} + \frac{p_y^2}{2m} + 2V(Y) + U(y) ,
\end{equation}
which omits the coupling term.  We can then compare the alternative evolutions with and without the coupling, by looking at how the overlap between the states changes for the two evolutions:
\begin{eqnarray}
\ket{\Psi(t)} &=& \exp( - iHt/\hbar )\ket{\Psi_0} , \nonumber\\
\ket{\Psi_d(t)} &=& \exp( - iH_dt/\hbar )\ket{\Psi_0} .
\end{eqnarray}
The rate of change of the overlap is given by
\begin{eqnarray}
\frac{d}{dt} \bracket{\Psi_d(t)}{\Psi(t)} &=& \frac{i}{\hbar} \bra{\Psi_d(t)} \left( H_d - H \right) \ket{\Psi(t)} \nonumber\\
&\approx& \frac{i}{\hbar} \bra{\Psi_d(t)} \left( -V''(Y)y^2 \right) \ket{\Psi(t)}.
\end{eqnarray}
Specializing to $t=0$, we get
\begin{equation}
\frac{d}{dt} \bracket{\Psi_d(t)}{\Psi(t)} \biggr|_{t=0} \approx -\frac{i}{\hbar} \expect{V''(Y)}_{\phi_0} \expect{y^2}_{\psi_0} .
\end{equation}

The magnitude of this rate has some properties that one would want in a measure of compositeness, but it is not perfect.  In particular, the rate of change of the overlap does not distinguish between physically significant changes in the state due to coupling, and physically meaningless changes in the global phase.  It also treats the internal and center-of-mass degrees of freedom symmetrically, while we are more interested in the question of how well the center-of-mass degree of freedom is approximated as a single indivisible particle.

This overlap does suggest a related approach, however, which can overcome these difficulties.  Let us define the state in the {\it interaction picture}:
\begin{equation}
\ket{\tilde\Psi(t)} \equiv \exp(iH_d t/\hbar) \ket{\Psi(t)} .
\end{equation}
We can then define the reduced density matrix for the center-of-mass in the interaction picture by taking a partial trace over the internal degrees of freedom:
\begin{equation}
\tilde\rho(t) \equiv \tr_{\rm int} \left\{ \ket{\tilde\Psi(t)}\bra{\tilde\Psi(t)} \right\} .
\end{equation}
To solve for the evolution of $\ket{\tilde\Psi(t)}$ or $\tilde\rho(t)$ in general requires the ability to solve the Schr\"odinger equation.  But at $t=0$ with an initial product state, the rate of change of $\tilde\rho(0)$ simplifies:
\begin{equation}
\frac{d\tilde\rho}{dt} \biggr|_{t=0} = - \frac{i}{\hbar} \left[ \hat{Q} , \ket{\phi_0}\bra{\phi_0} \right] ,
\end{equation}
where $\hat{Q}$ is an effective Hamiltonian:
\begin{eqnarray}
\hat{Q} &=& \tr_{\rm int} \bigl\{ \left(V(Y+y) + V(Y-y) - 2V(Y)\right) \nonumber\\
&& \times \left( I \otimes \ket{\psi_0}\bra{\psi_0}\right) \bigr\} \nonumber\\
&\approx& V''(Y)\expect{y^2}_{\psi_0} .
\end{eqnarray}

This evolution will produce a nontrivial change in $\tilde\rho$ if and only if $\ket{\phi_0}$ is not an eigenstate of $\hat{Q}$.  We can measure this by looking at the variance:
\begin{equation}
M_C = \sqrt{\expect{ \Delta\hat{Q}^2 }_{\phi_0}} = \sqrt{\expect{\hat{Q}^2}_{\phi_0} - \expect{\hat{Q}}_{\phi_0}^2} .
\end{equation}
This quantity $M_C$ is real and nonnegative, and is nonzero if and only if there is nontrivial coupling between the internal and center-of-mass degrees of freedom.  The value of $M_C$ depends on the nature of the potential $V(Y)$, and on the states $\ket{\phi_0}$ and $\ket{\psi_0}$.  The factor of $\expect{y^2}_{\psi_0}$ means that this quantity doesn't have a simple maximum (and the approximate expansion of $V(Y\pm y)$ to second order will not be valid if $y^2$ is too large), but it should have a minimum, which may or may not be nonzero, depending on the system.  It also suggests that the states which will show large effects of compositeness have both $y^2$ and $V''(Y)$ large, and are not eigenstates of $V''(Y)$.

Choosing $M_C$ to be the square root of the variance has a suggestive physical interpretation.  Since the operator $\hat{Q}$ is essentially an effective interaction Hamiltonian, we can define an energy-time uncertainty principle relating the variance of $\hat{Q}$ and the timescale $\Delta t$ of a state transition:
\[
\frac{2M_C}{\hbar} \le \frac{1}{\Delta t} .
\]
So we can interpret $M_C$ as proportional to the rate at which the interaction between the center-of-mass and internal degrees of freedom will cause the center-of-mass state to become orthogonal to the center-of-mass state {\it without} interaction.  We expect this measure to be nonzero whenever interaction with the internal degrees of freedom causes decoherence; but it may be nonzero even in cases with no decoherence, if the composite structure produces a change in the center-of-mass evolution.

The dependence of $\hat{Q}$ on the second derivative of the potential is at first a bit surprising.  (The second derivative is simply related to a quantity in mechanics known as the {\it jerk} or {\it jolt}.)  But a bit of reflection reveals that the leading order contribution to the coupling between the internal and center-of-mass degrees of freedom actually depends on the {\it third} derivative.  To see that, suppose we expand the potential $V(Y+y)+V(Y-y)$ about some point $Y_0$.  We can see how different terms in this expansion will affect the internal and external degrees of freedom:
\begin{itemize}
\item There is a set of terms involving only $Y$:  $2V(Y_0) + 2V'(Y_0)(Y-Y_0) + \cdots = 2V(Y)$.  These produce no coupling between the internal and center-of-mass degrees of freedom.
\item All terms that are linear in $y$ cancel out in the expansion, as do all terms of odd power in y.
\item There is a term $V''(Y_0)y^2$ that will produce a harmonic potential for the internal degree of freedom, but which is not coupled to the center of mass.  (If the internal potential $U(y)$ is also harmonic, this is just a constant frequency shift.)
\item The lowest order term including both the internal and center-of-mass variables is $V^{(3)}(Y_0)(Y-Y_0)y^2$. This term will produce a parametric driving of the internal degree of freedom dependent on the center-of-mass position, and a force on the center-of-mass that depends on the internal degree of freedom.
\end{itemize}
We see from this that to produce nontrivial coupling between the internal and center-of-mass degrees of freedom requires a nonvanishing third derivative. Without that, the coupling produces at most a constant frequency shift, and can affect only the global phase of the center-of-mass state.  The proposed measure $M_C$ is nonzero precisely when such nontrivial couplings are present.

A number of straightforward variants of this measure are possible.  For instance, we could relax the restriction to product states, or to pure states.  We could also look at the change over a finite time interval, rather than the instantaneous rate of change.  These differences would entail more computational complexity, but are worth exploring in future work.

\section{Solution for a Harmonic Internal Potential}

\subsection{Effective Hamiltonian}

Suppose the internal potential is harmonic, which is a good approximation for many simple molecules. 
\begin{equation}
U(x_1,x_2) = \frac{k}{2}(x_1 - x_2)^2 = 2ky^2  .
\end{equation}
The propagator becomes
\begin{eqnarray}
\label{eq:propagatorThree}
&& K(Y_f,y_f,t_f; Y_i,y_i,t_i) \nonumber\\
&=& \int\mathcal{D}Y
  e^{\frac{i}{\hbar}\int_{t_i}^{t_f} \left( \frac{M}{2} \dot{Y}^2(t) - 2V(Y(t)) \right) dt} \\
&&\times  \int\mathcal{D}y
  e^{\frac{i}{\hbar}\int_{t_i}^{t_f} \left( \frac{M}{2} \dot{y}^2(t) - (V''(Y(t))+2k)y^2(t) \right) dt} . \nonumber
\end{eqnarray}

For the purpose of evaluating the path integral over the internal degree of freedom, let us {\it assume that the center-of-mass trajectory} $Y(t)$ {\it is fixed}.  We can then define the time-dependent internal potential
\begin{equation}
W(y,t) \equiv \frac{w(t)}{2} y^2
\end{equation}
where
\begin{equation}
w(t) \equiv 4k + 2V''(Y(t)) .
\end{equation}
First let us set $M=1$ and $\hbar = 1$.  Let us make a change of variables:
\begin{eqnarray}
a \equiv \frac{1}{\sqrt2} (y + i p_y) , && y \equiv \frac{1}{\sqrt2} (a + a^\dagger) , \nonumber\\
a^\dagger \equiv \frac{1}{\sqrt2} (y - i p_y) , && p_y \equiv \frac{i}{\sqrt2} (a^\dagger - a) .
\end{eqnarray}
In terms of these new variables the effective Hamiltonian corresponding to the internal degree of freedom in (\ref{eq:propagatorThree}):
\begin{equation}
H_Y(t) = \Omega(t) a^\dagger a + \beta(t) (a^2 + {a^\dagger}^2) + \frac{1}{2} \Omega(t) ,
\end{equation}
where
\begin{eqnarray}
\Omega(t) &\equiv& \frac{w(t) + 1}{2} = \frac{2V''(Y(t)) + 4k + 1}{2} , \\
\beta(t) &\equiv& \frac{\Omega(t) - 1}{2} = \frac{2V''(Y(t)) + 4k - 1}{4} .
\end{eqnarray}
Employing this Hamiltonian, for each trajectory in the center-of-mass path integral in (\ref{eq:propagatorThree}), we can reduce the problem of determining the dynamics of the internal degree of freedom, $y$, to the solution of a single ordinary (nonlinear) differential equation. 

\subsection{Solution for Fixed Center-of-Mass Path}

To solve for the dynamics, we begin by noting that operators can be put in normal-ordered form \cite{Louiselle73}:
\begin{equation}
f(a,a^\dagger) = \sum_{r,s\ge0} f_{r,s} (a^\dagger)^r a^s ,
\end{equation}
where the coefficients $f_{r,s}$ are complex numbers.  Every such operator is therefore in one-to-one correspondence with a complex function $\bar{f}(\alpha,\alpha^*)$, where
\begin{equation}
\bar{f}(\alpha,\alpha^*) = \sum_{r,s\ge0} f_{r,s} (\alpha^*)^r \alpha^s
  = \bra{\alpha} f(a,a^\dagger)  \ket{\alpha} .
\end{equation}
We can now use the fact that the function $\bar{f}(\alpha,\alpha^*)$ determines the operator $f(a,a^\dagger)$ to solve for the unitary propagator of the parametrically-driven oscillator.

Let $U(t)$ be the unitary propagator from the initial time to time $t$.  It satisfies the operator Schr\"odinger equation:
\begin{eqnarray}
i\frac{dU}{dt} (t) &=& H(t)U(t) \\
&=& \left[ \Omega(t) a^\dagger a + \beta(t) (a^2 + {a^\dagger}^2) + \frac{1}{2} \Omega(t)\right] U(t) . \nonumber
\end{eqnarray}
We can write $U$ in normal-ordered form,
\begin{eqnarray}
U &=& \sum_{n,m} U_{n,m} (a^\dagger)^n a^m \Rightarrow \nonumber\\
\bar{U}(\alpha,\alpha^*) &=& \sum_{n,m} U_{n,m} (\alpha^*)^n \alpha^m
  = \bra{\alpha}U\ket{\alpha} .
\end{eqnarray}
This function obeys a differential equation:
\begin{eqnarray}
&& i\frac{d\bar{U}}{dt}(\alpha,\alpha^*,t) = \bra{\alpha}H(t) U(t) \ket{\alpha} \\
&=& \bra{\alpha}\left[ \Omega(t) a^\dagger a + \beta(t) (a^2 + {a^\dagger}^2) + \frac{1}{2} \Omega(t)\right] U(t)\ket{\alpha} . \nonumber
\end{eqnarray}
Employing the simple results:
\[
\bra{\alpha} (a^\dagger)^n U(t) \ket{\alpha} = (\alpha^*)^n \bar{U}(t) ,
\]
\[
\bra{\alpha} a^m U(t) \ket{\alpha} = (\alpha + \partial/\partial\alpha^*)^m \bar{U}(t) 
\]
the differential equation for $\bar{U}(t)$ becomes
\begin{eqnarray}
&&i\frac{d\bar{U}}{dt} (\alpha,\alpha^*,t) = \biggl[ \Omega(t) \alpha^* (\alpha  + \partial/\partial\alpha^*) \\
&+& \beta(t) \left((\alpha  + \partial/\partial\alpha^*)^2 + (\alpha^*)^2 \right)
  + \frac{1}{2} \Omega(t) \biggr] \bar{U}(\alpha,\alpha^*,t) . \nonumber
\label{eq:uBarDiffEq}
\end{eqnarray}

We make an ansatz for the solution:
\[
\bar{U}(\alpha,\alpha^*,t) = e^{-\frac{i}{2}\int_0^t \Omega(t') dt'} e^{G(\alpha,\alpha^*,t)} ,
\]
where
\begin{equation}
G(\alpha,\alpha^*,t) = A(t) + D(t) \alpha^*\alpha + E(t) {\alpha^*}^2 + F(t) \alpha .
\label{eq:ansatz}
\end{equation}
If we can solve for the functions of time in (\ref{eq:ansatz}), we can easily recover the effective propagator for the internal states:
\begin{eqnarray}
&&\int_{y_i}^{y_f} \mathcal{D}y e^{\frac{i}{\hbar} \int_{t_i}^{t_f} \left( \frac{M}{2} \dot{y}^2(t)
  - \left( V''(Y(t)) + 2k \right)y^2(t) \right) dt} \\
&=& e^{-\frac{i}{\hbar} \int_0^t \Omega(t') dt'} e^{A(t)} \bra{y_f}
  e^{D(t) a^\dagger a + E(t) \left(a^dagger\right)^2 + F(t) a} \ket{y_i} . \nonumber
\end{eqnarray}
Note that one might expect also terms proportional to  $\alpha^*$ and $\alpha^2$ but these terms can be chosen to be identically zero.  

At $t=0$ we want $\bar{U} = 1$, which gives us initial conditions:
\[
A(0) = D(0) = E(0) = F(0) = 0 .
\]
Plugging the above expression into the differential equation (\ref{eq:uBarDiffEq}) yields the following differential equations for the functions $A(t)$, $D(t)$, $E(t)$ and $F(t)$:
\begin{eqnarray}
i \frac{dA}{dt} &=& 2\beta(t) E(t) , \nonumber\\
i \frac{dD}{dt} &=& \left[ \Omega(t) + 4 \beta(t)E(t) \right] (D(t) + 1) , \\
i \frac{dE}{dt} &=& \beta(t) + 2 \Omega(t) E(t) + 4\beta(t) E^2(t) , \nonumber\\
i \frac{dF}{dt} &=& \beta(t)  (D(t) + 1)^2 . \nonumber
\end{eqnarray}
Looking at the structure of these equations, we see that one can solve for $A(t)$, $D(t)$ and $F(t)$ in terms of $E(t)$:
\begin{equation}
A(t) = -2i \int_0^t \beta(t') E(t') dt' ,
\end{equation}
\begin{equation}
D(t) = e^{-i \int_0^t (\Omega(t') + 4 \beta(t') E(t')) dt' } - 1,
\end{equation}
\begin{equation}
F(t) = -i \int_0^t \beta(t') e^{-2i \int_0^{t'} (\Omega(t'') + 4 \beta(t'') E(t')) dt'' } dt' .
\end{equation}
The equation for $E(t)$ is self-contained, and must be solved before plugging it into these expressions:
\begin{equation}
i \frac{dE}{dt} = \beta(t) + 2 \Omega(t) E(t) + 4\beta(t) E^2(t) .
\end{equation}
For certain choices of $w(t)$ this equation may be solvable in closed form; but generically, this equation may only be solvable numerically.  This would be sufficient to allow the solution of the influence functional (\ref{eq:propagatorTwo}) for particular choices of $Y(t)$ and $Y'(t)$. It is not practical to evaluate the path integrals in (\ref{eq:transitionProbApprox}) in this way, however it is often the case that a path integral is dominated by the classical path \cite{Feynman65}. In that case, by replacing $Y(t)$ and $Y'(t)$ in $\bracket{{\psi}_Y}{{\psi}_{Y'}}$ by $Y_{cl}(t)$ and $Y'_{cl}(t)$, the leading order to  (\ref{eq:transitionProbApprox}) could be calculated.

\section{Example of interference suppression}

We can see from the above results that coupling to internal degrees of freedom should generically lead to suppression of interference effects, just like other kinds of decoherence.  It is hard to estimate the magnitude of this effect analytically, however.  In this section, we give a numerical analysis of interference in a simple scattering process, and show that even a single internal degree of freedom can significantly suppress interference.

Suppose two wave packets, initially separated from each other, approach a potential of compact support, such as a square well, from opposite directions. Each wave packet, taken alone, would produce both a reflected and transmitted wave. By linearity, when we start with the superposition of the two, each will produce such waves, and they will interfere in some manner, depending on the initial conditions. This interference is a strictly quantum effect with no classical counterpart. We show below that, when the scattered particles possess internal states, as the entanglement between internal state and environment grows, the interference diminishes.

\subsection{The scattering experiment}

Consider a particle moving in one dimension with a potential $V(x)$ which is zero everywhere outside of a compact region of size L: $-L/2 \le x \le L/2$. For simplicity, we will assume a simple square well potential:
\begin{equation}
V(x) = \left\{ \begin{array}{cc} 0 , & x<-L/2 ,\\
  -V_0 , & -L/2 \le x \le L/2 , \\
  0 , & x > L/2 , \end{array} \right.
\end{equation}
where $V_0 > 0$.  A narrow wave packet incident on this potential will generically scatter into two transmitted and reflected wave packets. Classically, incident particles would not be reflected from such a well; such reflection is already a quantum-mechanical effect.  

\begin{figure}[htbp]
\begin{center}
\includegraphics[width=3in]{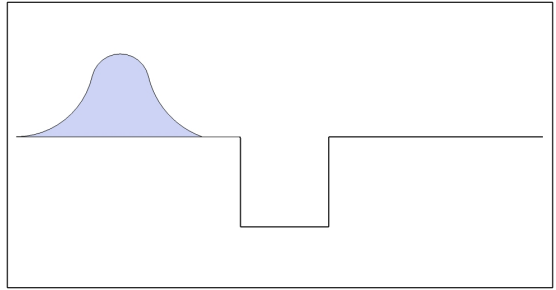}
\caption{(Color online.)  A quantum particle scattering from a square well.}
\label{fig:scattering}
\end{center}
\end{figure}

Suppose we have a particle of mass $m=1$ incident from the left in a plane wave state with momentum $p$; then it is straightforward to calculate the norms of the reflection and transmission coefficients $R$ and $T$:
\begin{eqnarray}
|R|^2 &=& \frac{2(mV_0)^2 \sin^2(L\tilde{p})}{2p^4 + 4(mV_0)p^2 + 2(mV_0)^2\sin^2(L\tilde{p})} , \\
|T|^2 &=& \frac{2p^2(p^2+2mV_0)}{2p^4 + 4(mV_0)p^2 + 2(mV_0)^2\sin^2(L\tilde{p})} ,
\end{eqnarray}
where
\[
\tilde{p} = \sqrt{p^2 + 2mV_0} .
\]

By choosing an appropriate set of values $p$, $L$  and $mV_0$ it is possible to have this square well act like a 50/50 beam splitter for wave-packets centered around momentum $p$.  If we choose $p=1$ and $L=0.5$ in dimensionless units, this 50/50 reflection/transmission occurs for $mV_0 \approx 2.64$.

We can then model a situation like the second half of a Michelson interferometer:  a superposition of two identical wave packets incident from the right and left with the same magnitude of $|k|$ can interfere almost perfectly, so that a single wavefunction emerges either to the right or to the left (depending on the relative phase of the two incident wave packets).

\subsection{Scattering of a composite particle}

Now suppose that instead of a single particle of mass $m=1$ we have a composite particle, comprising two particles of mass $1/2$ bound by a harmonic potential.  Because the potential is piecewise constant, $V''(x) = 0$ everywhere except at $x = \pm L/2$.  So from the results of the previous section, the internal and center-of-mass degrees of freedom are decoupled everywhere except at those two points, where there is an impulsive coupling between them.  

If we prepare the center-of-mass degree of freedom in a narrow wave packet and scatter it off of this potential well, entanglement can be produced between the internal and center-of-mass degrees of freedom. Whether entanglement is in fact produced depends on the initial state of the internal degree of freedom. In a classical harmonic oscillator, the strength of parametric driving depends on the initial amplitude of oscillation. An oscillator in equilibrium experiences no driving. Similarly, if the internal degree of freedom begins in the ground state, the parametric driving from the center-of-mass degree of freedom will be ineffective in coupling the two degrees of freedom.  If the internal degree of freedom begins in an excited state, however, it will be more strongly coupled to the center-of-mass degree of freedom.

\subsection{Numerical results}

\begin{figure}[htbp]
\begin{center}
\includegraphics[width=3in]{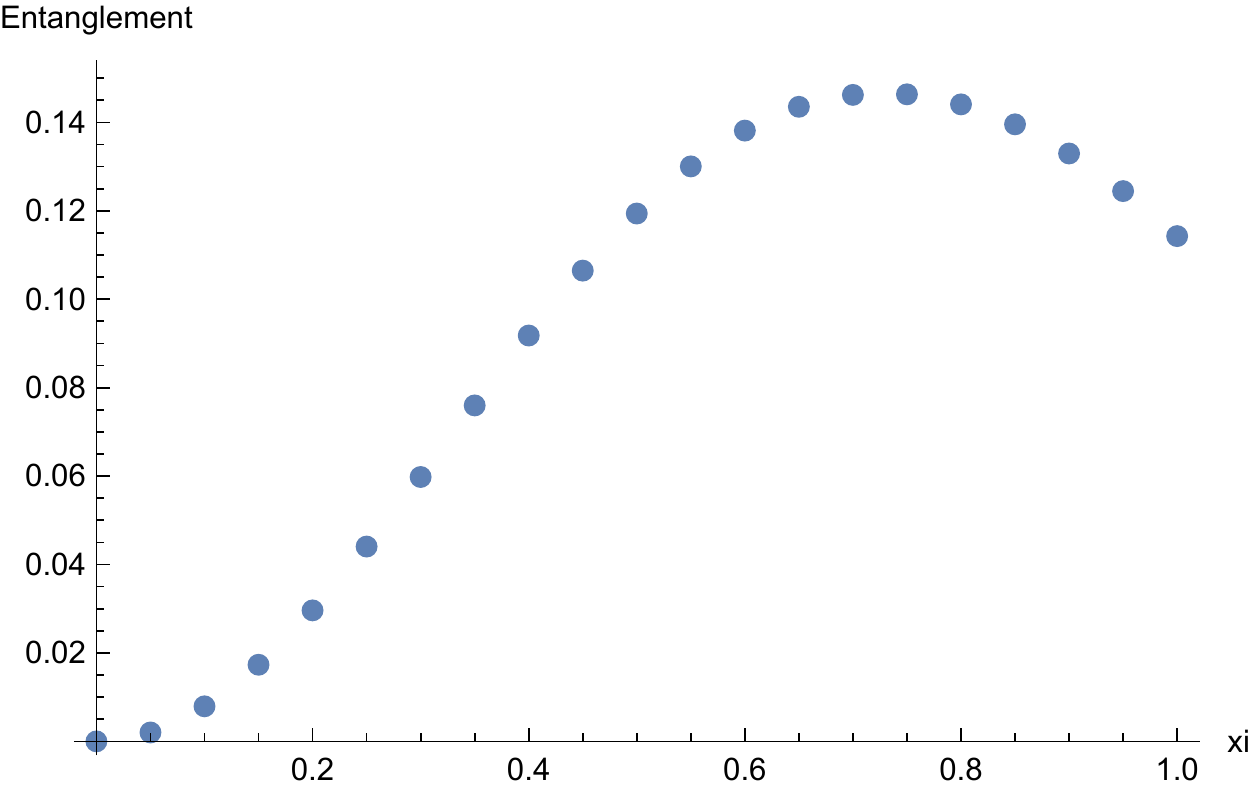}
\caption{(Color online.)  Entanglement between the internal and center-of-mass degrees of freedom, vs. the initial excitation of the internal degree of freedom.  The measure is the purity of the reduced density matrix.}
\label{fig:entanglement}
\end{center}
\end{figure}

\begin{figure}[htbp]
\begin{center}
\includegraphics[width=3in]{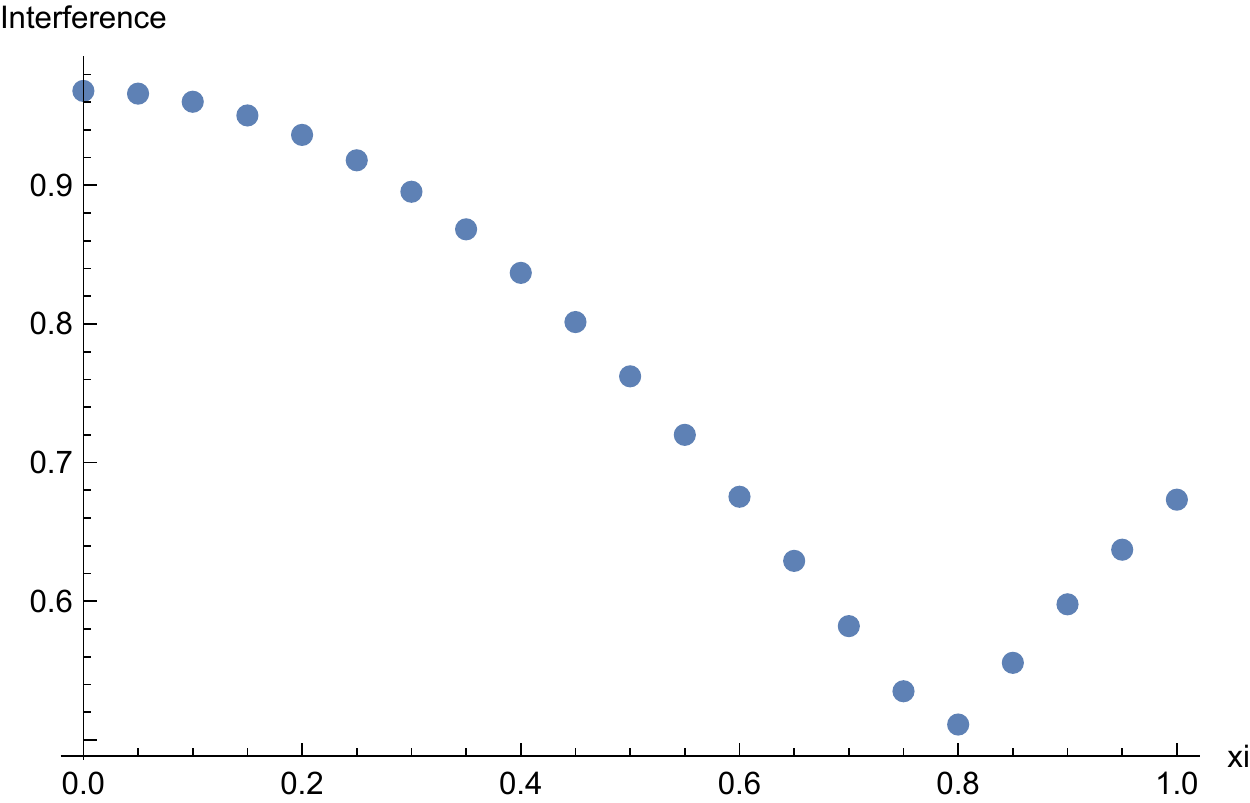}
\caption{(Color online.)  Maximum interference between left-going and right-going Gaussian wave packets scattering off the central potential well, vs. the initial excitation of the internal degree of freedom.  Initially the internal and center-of-mass degrees of freedom are in a pure product state.}
\label{fig:interference}
\end{center}
\end{figure}

We have numerically simulated the interference experiment described in the previous subsection for two coupled particles with a variety of initial states.  The center-of-mass degree of freedom begins in the superposition state
\begin{eqnarray}
\phi_i(Y) &=& \frac{1}{{\left(4\sigma_y^2 \pi\right)}^{1/4}} \biggl( \exp\left(-(Y+Y_0)^2/2\sigma_Y^2 + ipY\right) \nonumber\\
&&  + e^{i\theta} \exp\left(-(Y-Y_0)^2/2\sigma_Y^2 - ipY\right) \biggr),
\end{eqnarray}
and the internal degree of freedom begins in the state
\begin{equation}
\psi_i(y) = \frac{1}{{\left(\sigma_y^2 \pi\right)}^{1/4}} \exp\left(-(y-\xi_0)^2/2\sigma_y^2 \right) ,
\end{equation}
where $Y_0 = 20$, $\sigma_Y=5$, $p=1$, $L=0.5$, and $mV_0 = 2.64$.  

The parameter $\theta$ is a relative phase between the two wave packets; we choose this phase $\theta$ to maximize the interference between them.  We choose the internal coupling strength $k=(9/2)^2$, and the width of the initial wave packet for the internal degree of freedom to be $\sigma_y = 1/9$; with this choice, the initial state $\psi_i(y)$ is the ground state for $\xi_0=0$, and an excited state for $\xi_0\ne0$.  We have simulated this system from its initial state until after the scattering (into two wave packets) from the potential well for a range of values of $\xi_0$.

Decoherence generally has its effect by suppressing interference.  In this case, the interference is between the reflected and scattered components of the two incoming wave packets.  We choose the relative phase $\theta$ to maximize the probability after scattering that the scattered particle will be moving {\it leftward}.  (Of course, the situation would be completely symmetric if we chose to maximize the outgoing rightward component.)

Since the state of this system is pure at all times, the proper measure of entanglement between the internal and center-of-mass degrees of freedom would be the {\it entropy of entanglement} between them.  However, this is a somewhat messy quantity to calculate for high-dimensional systems like those we are treating here.  Instead, we use a simpler proxy quantity:  the {\it impurity} of the reduced density matrix of the internal degree of freedom.  This is simply
\[
E = 1 - \tr\{ \rho_{\rm int}^2 \} ,\ \ \ \rho_{\rm int} = \tr_{\rm com}\{ \ket{\Psi}\bra{\Psi} \} .
\]

Results of the simulations are plotted in Figs.~\ref{fig:entanglement} and \ref{fig:interference}.  Fig.~\ref{fig:entanglement} plots the impurity of the reduced density matrix for the internal degree of freedom after scattering as a function of its initial displacement.  Fig.~\ref{fig:interference} plots the maximum leftward probability of the scattered particle.  

We see that if the internal degree of freedom begins in the ground state, the parametric driving is ineffective in coupling the internal and center-of-mass degrees of freedom, so the two degrees of freedom remain unentangled and the interference is maximal. This is the case for simple molecules like water, in which $\hbar\omega$ is about 18 times $kT$ at room temperature \cite{Miller01}.

As we increase the displacement of the initial state of the internal degree of freedom, the state of the two degrees of freedom becomes entangled, and the interference is reduced.  Increasing the displacement increases the entanglement and decreases the interference, up to a point where the two curves turn over and the entanglement diminishes.

This turn over is not too surprising.  Coupling to a single degree of freedom can only modestly suppress interference, in general; such small systems commonly exhibit recurrences and ``re-coherences.'' Models of environmental decoherence that produce monotonic decay of interference typically assume large heat baths with many degrees of freedom.  Such an effect could probably be caused by a large composite system with many internal degrees of freedom, provided that the center-of-mass coupled strongly enough to these internal modes.  But the difficulties of solving such a system analytically are great, and they quickly become intractable numerically as well. However, even in this very simple system, we see a very close relationship between the degree of entanglement and the suppression of interference.  This relationship makes intuitive sense, based on earlier experience with external environments---the state of the internal degree of freedom records ``which path'' information about the center-of-mass, specifically, information about whether it was transmitted or reflected by the potential well.

\subsection{Measure of compositeness}

What about the measure of compositeness defined in Section \ref{sec:compositeness}?  We can calculate this numerically for the scattering system described in this section.  We use the same potential as in the simulations, and calculate
\[
M_C = \sqrt{ \expect{ \Delta\hat{Q}^2 }_{\phi} } ,
\]
where
\[
\hat{Q} = \tr\left\{ \left( 2V(Y)-V(Y+y)-V(Y-y)\right) \left( I\otimes \ket{\psi}\bra{\psi} \right) \right\} .
\]
We choose states $\ket{\phi}$ for the center-of-mass and $\ket{\psi}$ for the internal degree of freedom of the same form as used in the numerical simulations,
\[
\phi(Y) = \frac{1}{{\left(4\sigma_y^2 \pi\right)}^{1/4}} \exp\left(-(Y-Y_0)^2/2\sigma_Y^2 \right) ,
\]
\[
\psi_i(y) = \frac{1}{{\left(\sigma_y^2 \pi\right)}^{1/4}} \exp\left(-(y-\xi_0)^2/2\sigma_y^2 \right) ,
\]
with the two wavepackets centered at $Y_0$ and $\xi_0$, respectively.  We can then calculate $M_C$ for different choices of $Y_0$ and $\xi_0$.

\begin{figure}[htbp]
\begin{center}
\includegraphics[width=3in]{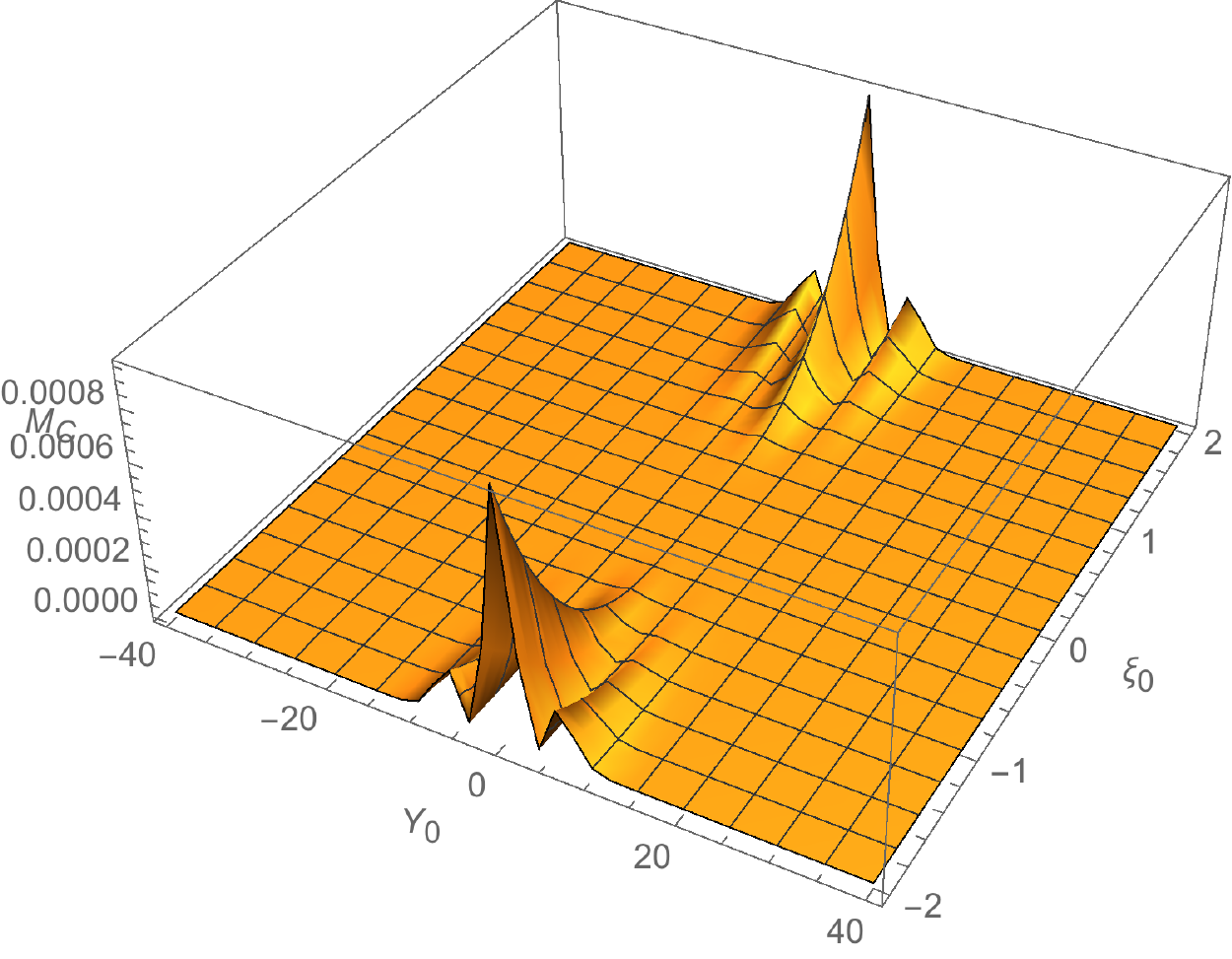}
\caption{(Color online.)  Expectation $\expect{\hat{Q}}^2$ for wavepacket centered at $Y_0$ (center-of-mass) and $\xi$ (internal).  The parameters in this plot are the same as in the numerical simulations.}
\label{fig:expectQ}
\end{center}
\end{figure}

\begin{figure}[htbp]
\begin{center}
\includegraphics[width=3in]{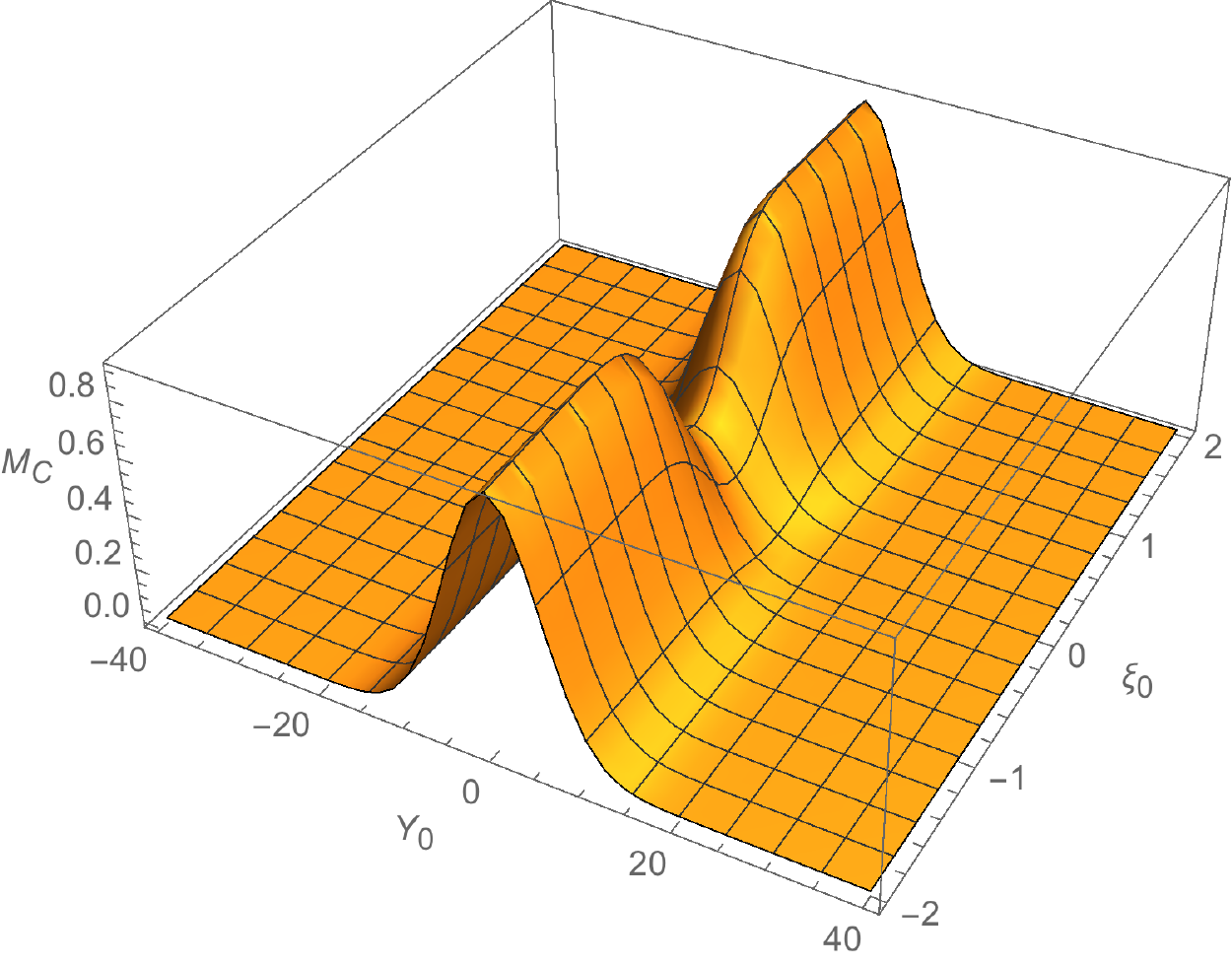}
\caption{(Color online.)  Compositeness $M_C = \sqrt{\expect{\Delta\hat{Q}^2}}$ for wavepacket centered at $Y_0$ (center-of-mass) and $\xi$ (internal).  The parameters in this plot are the same as in the numerical simulations.  Because the potential well is narrow compared to the width of the center-of-mass wavepacket, the composite system scatters from the potential as a whole.}
\label{fig:delQ2a}
\end{center}
\end{figure}

Figures \ref{fig:expectQ} and \ref{fig:delQ2a} show the expectation $\expect{\hat{Q}}^2$ and the compositeness measure $M_C = \sqrt{\expect{\Delta\hat{Q}^2}}$ for the same parameters used in the numerical simulations above.  While in the expectation $\expect{\hat{Q}}^2$ we can see the effects of the two potential well edges, the width of the wavepacket means that the composite system essentially scatters off of the potential well as a whole.  From Fig.~\ref{fig:delQ2a} we see that the evolution is significantly affected by the coupling between the internal and center-of-mass degrees of freedom throughout the scattering region.

\begin{figure}[htbp]
\begin{center}
\includegraphics[width=3in]{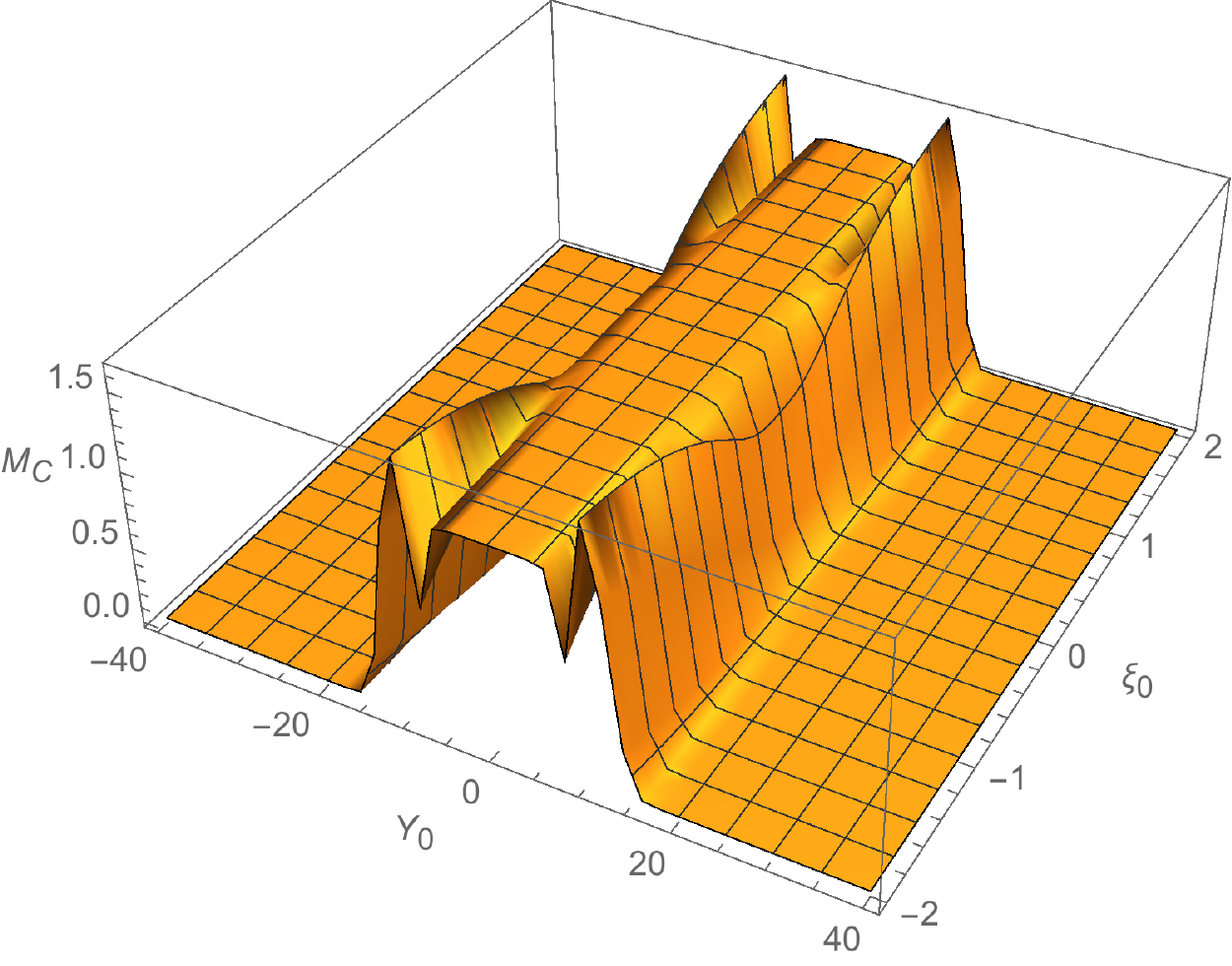}
\caption{(Color online.)  Compositeness $M_C = \sqrt{\expect{\Delta\hat{Q}^2}}$ for wavepacket centered at $Y_0$ (center-of-mass) and $\xi$ (internal).  Here we have used a narrower wavepacket $\sigma_y^2 = 1$ and a wider potential well $L=20$ so that we can see coupling due to separate scatterings from the edges of the potential well, as well as from the potential as a whole.}
\label{fig:delQ2b}
\end{center}
\end{figure}

For a narrow wave packet and/or a broader potential well, the coupling should reflect separate scatterings off of the two edges of the potential.  Indeed, plotting $M_C$ for a case like that demonstrates the features of these two separate scatterings.  In both these cases, we see that the coupling between internal and center-of-mass degrees of freedom leads to significant differences from the evolution of a single, indivisible particle.

\section{Conclusions}

As we have shown in this paper, it is possible for coupling between the center-of-mass and internal degrees of freedom in a composite object to give rise to decoherence.  We have calculated expressions for coupling to a single internal vibrational mode, and numerically demonstrated that this can lead to suppression of interference in a simple scattering experiment.  Just as in coupling to an external environment, a signature of this decoherence is the growth of entanglement between the center-of-mass and internal degrees of freedom.  For this work, the overall state of the system was pure, so the impurity of the reduced density matrix for the internal degree of freedom served as an easy-to-calculate proxy for the entropy of entanglement.

However, there are certain important differences between decoherence due to internal degrees of freedom and due to an external environment.  The coupling between the internal degrees of freedom and the center-of-mass takes the form of a parametric driving term, and is dependent on the second derivative of the background potential.  Therefore we would only expect to see such an effect when the background potential is suitably nontrivial, and only when the internal degrees of freedom are in an appropriate initial state.  For internal degrees of freedom in an initial ground state, the effect of parametric driving is weak or nonexistent.

This raises the intriguing question of whether it would be possible to see this effect experimentally in a controllable fashion.  Matter interferometry experiments have been done with molecules, including large molecules with many internal degrees of freedom.  It might be possible to vary the coupling between the internal and center-of-mass degrees of freedom by exciting one or more vibrational modes of the molecules before the interference experiment, and observing the effect on the visibility of the interference.  For three-dimensional systems, there can also be potential couplings with rotational degrees of freedom \cite{Hillery05}.

A number of interesting questions remain unanswered.  A single internal mode cannot really act as a decohering bath.  It might be possible to derive effects for a composite particle with many internal degrees of freedom.  In this paper we have suggested one possible measure of compositeness---that is, the degree to which we can approximate a composite system as a single particle without internal structure.  This quantity is based on the coupling between the internal and center-of-mass degrees of freedom, and how large an effect this has on the reduced state of the center-of-mass degrees of freedom.  But a broader exploration of how well this measure correlates with (for example) suppression of interference by internal degrees of freedom has yet to be made.  This measure could also be varied in a number of ways.  These questions should be grounds for interesting future work.

\section*{Acknowledgments}

LM and TAB would like to acknowledges Mark Hillery, Jos\'e Gonzalez, Kung-Chuang Hsu, Jan Florjanczyk, and Christopher Cantwell for interesting and useful conversations.  They also acknowledge the support of Caltech's Institute for Quantum Information and Matter (IQIM).

\end{document}